\documentclass[a4paper]{iopart}

\usepackage{graphicx}
\usepackage[subrefformat=parens,labelformat=parens]{subfig}
\usepackage{color}
\usepackage{xspace}
\usepackage{url}
\usepackage{booktabs} 

\usepackage{threeparttable}

\usepackage{lineno} 

\usepackage{hyperref}
\definecolor{linkcolor}{rgb}{0,0,1}
\definecolor{urlcolor}{rgb}{0,0,.7}
\definecolor{citecolor}{rgb}{0,0,1}
\definecolor{acrocolor}{rgb}{0,0,.7}
\hypersetup{bookmarksopen,colorlinks=true}
\hypersetup{pdfstartview=FitH}
\hypersetup{linktocpage=true,bookmarksnumbered=true}
\hypersetup{plainpages=false,breaklinks=true}
\hypersetup{linkcolor=linkcolor,citecolor=citecolor,urlcolor=urlcolor}

\def\aligned{\vcenter\bgroup\let\\\cr
\halign\bgroup&\hfil${}##{}$&${}##{}$\hfil\cr}
\def\endaligned{\crcr\egroup\egroup}

\begin{document}
\title{Thermal modelling of Advanced LIGO test masses}
\author{H. Wang$^1$, C. Blair$^2$, M. Dovale \'Alvarez$^1$, A. Brooks$^3$, M. F. Kasprzack$^4$, J. Ramette$^5$, P. M. Meyers$^6$, S. Kaufer$^7$, B. O'Reilly$^8$, C. M. Mow-Lowry$^1$, A. Freise$^1$}
\address{$^1$ University of Birmingham, Birmingham, B15 2TT, UK}
\address{$^2$ University of Western Australia, Crawley, WA 6009, Australia}
\address{$^3$ California Institute of Technology, Pasadena, CA 91125, USA}
\address{$^4$ Louisiana State University, Baton Rouge, LA 70803, USA}
\address{$^5$ Hillsdale College, Hillsdale, Michigan 49242, USA}
\address{$^6$ University of Minnesota, Minneapolis, Minnesota 55455, USA}
\address{$^7$ Leibniz University of Hannover, D-30167 Hannover, Germany}
\address{$^8$ LIGO Livingston Observatory, Livingston, LA 70754, USA}
 \ead{haoyu@star.sr.bham.ac.uk}
 
\begin{abstract}
High-reflectivity fused silica mirrors are at the epicentre of today's advanced gravitational wave detectors. In these detectors, the mirrors interact with high power laser beams.  As a result of finite absorption in the high reflectivity coatings the mirrors suffer from a variety of thermal effects that impact on the detectors' performance. We propose a model of the Advanced LIGO mirrors that introduces an empirical term to account for the radiative heat transfer between the mirror and its surroundings. The mechanical mode frequency is used as a probe for the overall temperature of the mirror. The thermal transient after power build-up in the optical cavities is used to refine and test the model. The model provides a coating absorption estimate of 1.5 to 2.0\,ppm and estimates that 0.3 to 1.3\,ppm of the circulating light is scattered on to the ring heater.


\end{abstract}


\section{Introduction: \normalfont{\textit{How thermal transients affect Advanced LIGO}}}


Advanced LIGO is currently the most sensitive length sensing device ever created. The first gravitational wave signal ever detected was recorded in Advanced LIGO's first observation run (O1) by both the Hanford and Livingston observatories~\cite{Observation15}, immediately after the completion of upgrades from initial LIGO. 
The upgraded detectors, which consist of Michelson interferometers with 4\,km Fabry-Perot optical cavities as arms, use high laser power to increase their sensitivity~\cite{AdvancedLIGO15}. 
Each arm cavity is composed of fused silica mirrors with high-reflectivity coatings, known as the input test mass (ITM) and end test mass (ETM). The optical power in each cavity will eventually reach 800\,kW~\cite{Harry10}.  
A small portion of this power will be absorbed by test masses and converted into heat. 
When the arm cavities control systems are locked, resulting in the optical power build-up, the absorbed heat creates a thermal transient in the mirrors~\cite{Winkler91}.  
This thermal transient impacts the performance of the interferometer in many ways while the two following effects currently dominate the interferometer's behaviours:


1. Thermal lensing~\cite{Strain94, Degallaix04, Zhao2006} is caused by a change of refractive index via the thermo-optic effect. It induces a change in the optical path length within optical components resulting in aberrations in the power recycling and signal recycling cavities~\cite{AdvancedLIGO15}. Such aberrations normally contribute to a deterioration of mode
matching~\cite{Rocchi11} and can reduce the optical gain, causing a reduced sensitivity of the interferometer.


2. A change in the tuning conditions for parametric instabilities~\cite{Braginsky01, Evans15, Blair16} is caused by two factors. 
First, thermal expansion deforms the high-reflectivity coatings of the test masses. The first order deformation is a change of the radius of curvature (ROC) of the mirror~\cite{Hello1990}, which will shift the frequency of the transverse optical modes (TEM) resonant in the cavity, changing the mode spacing between the fundamental mode and TEM modes (by $\sim10^{3}$\,Hz). Second, the Young's modulus of the mirror substrates has a small positive thermal dependence which results in an increase ($\sim10^{-1}$\,Hz) in the mechanical mode frequencies as the mirror warms. Parametric instabilities are most severe when the mechanical mode frequency equals the frequency spacing between the fundamental mode and TEM modes, a condition that is altered by the thermal change, where the interferometer may become unstable and inoperable.


These issues need to be addressed in order to minimise their impact on detector sensitivity. 
The power of the circulating beam in arm cavities at Advanced LIGO during O1 was 100\,kW and thermal effects were managed with various mitigation strategies~\cite{Brooks16}. 
However, these effects will become more severe in future  as the detectors employ higher laser power to reach design sensitivity. 
Many of the current mitigation strategies will therefore require further attention.

To aid the design and development of such strategies, a good thermal model of test masses, along with good estimates of their coating absorptions, is required. 
Monitoring the coating absorption also provides a means of identifying coating damage or contamination. The coating absorption of Livingston's Y-arm end test mass (ETMY) was measured to be less than 0.5\,ppm~\cite{optics, Billingsley12} before installation. New on site thermal lensing measurements recently showed that the coating absorption of ETMY is 2.1\,ppm~\cite{HWS}. The increased coating absorption may be indicative of such contamination.


In this paper we build a thermal model of the test mass and its surroundings that uses shifts in mechanical mode frequencies as a probe for the overall temperature of the mirror. 
Thermal transients on the timescale of minutes to a few hours are well understood, which enables us to estimate the energy absorbed by the ETMY coating. Scattered light heating surrounding elements is observed and introduced in the model. A model of the long term thermal behaviour of the test mass is then introduced, where complex structures in the vicinity of the optic are simplified to a single element. Such an empirical model helps predict the thermal behaviour on the timescale of tens of hours.

This model provides good estimates of the coating absorption of the ETMY of the Livingston detector. Using the model for ongoing absorption monitoring provides early warning for potential contamination or degradation of test masses and tracking mechanical mode frequencies that might be relevant for parametric instability mitigation strategies. 

\section{Method: \normalfont{\textit{Using acoustic modes as temperature probes}}} \label{sec:Method}
Mechanical mode frequencies can be used as very accurate test mass thermometers. This technique was first introduced by M. Punturo and F. Travasso in 2001~\cite{Day10} as a method to characterize the mirror absorption of the French-Italian gravitational wave detector Virgo~\cite{caron97, Accadia:2012zzb, TheVirgo:2014hva}. Mechanical mode frequencies depend on the mirror dimensions and on two material properties that have a temperature dependence: the Young's modulus and the Poisson ratio. The eigenfrequencies of a cylinder can be expressed analytically as~\cite{Chree1886, Lamb1917, Strigin2008}
\begin{equation}
\omega_{m} = \beta_m \sqrt{\frac{E}{\rho(1+\nu)}},
\label{eq:modeFReq}
\end{equation}
where $\beta_m$ is a parameter encompassing several aspects of cylinder dimensions and has restricted values, $E$ is the Young's Modulus, $\nu$ is the Poisson ratio, and $\rho$ is the density of the material.


Our model shows that the change in eigenfrequencies due to the thermal expansion of the substrate is negligible when compared to the change due to the temperature dependence of $E$ and $\nu$. Most materials soften with increasing temperature, but fused silica's Young's modulus increases in the $(-200,\,1100)\,^{\circ}$C interval~\cite{Spinner1956, Spinner1960} and the rate of change at $17~^{\circ}$C is 11.5\,MPa/K. The Poisson ratio of fused silica also increases in the $(0,\,1200)\,^{\circ}$C interval with a rate of $5.5\times 10^{-5}/$K at $17\,^{\circ}$C~\cite{Spinner1956, PoissonRatio}. These rates can be considered constant within the small temperature range considered here of $\pm 0.6\,^{\circ}$C around $17\,^{\circ}$C. Equation~\ref{eq:modeFReq} implies that temperature-induced changes of $E$ and $\nu$ have contributions with opposite sign to the mechanical mode frequencies with a ratio of $3.4:-1$. Therefore, the change in the mechanical mode frequencies due to the change in Young's modulus dominates over the change due to the Poisson ratio, and the eigenfrequencies increase as the test mass becomes warmer.

\begin{figure}[!ht]
\begin{center}
\includegraphics[scale=0.6]{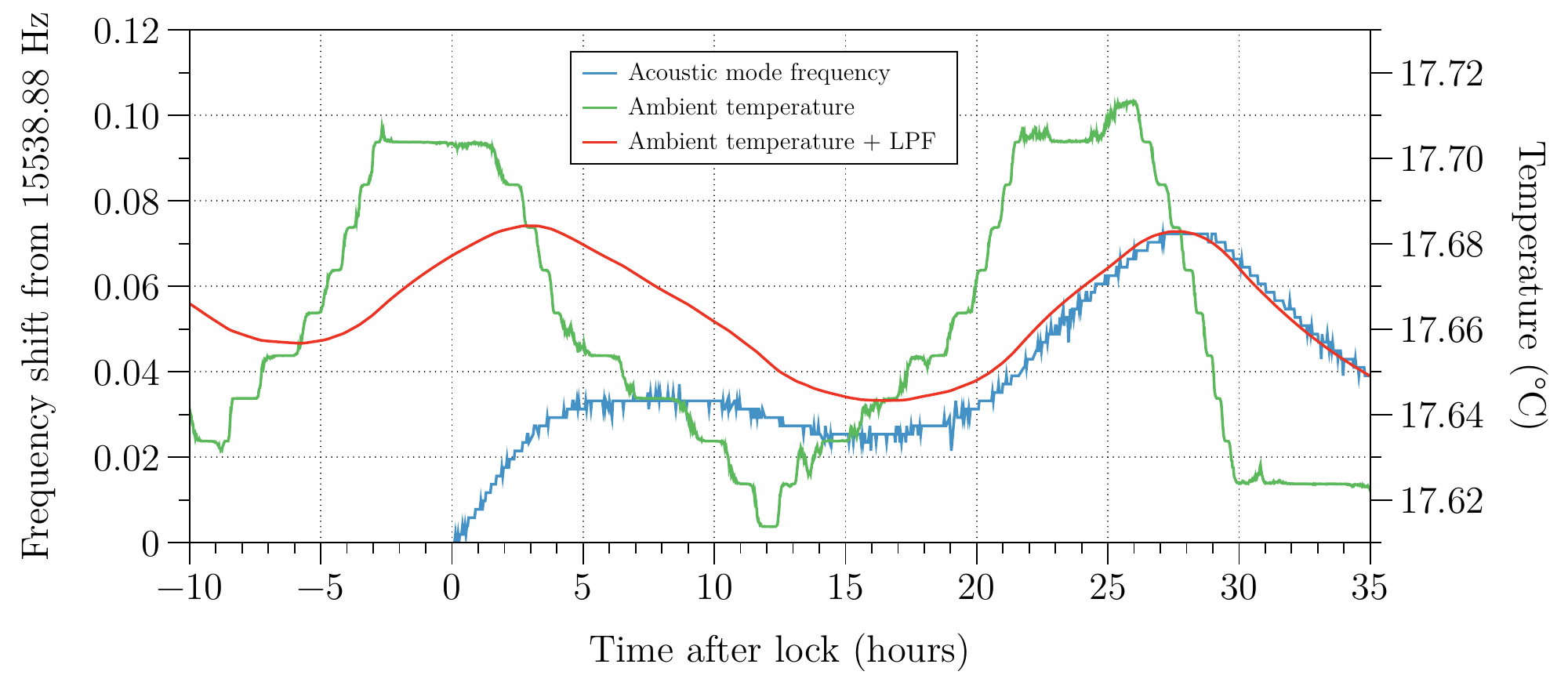}
\caption{Frequency shift of the 15540\,Hz eigenmode of LIGO Livingston's ETMY and ambient temperature fluctuations during a 36\,hour lock.}
\label{fig:ModeAmb1008}
\end{center}
\end{figure}

Mechanical modes of test masses are sensed at the cavities' transmission port when the cavities are locked, and we experimentally track the frequency of the 15.5\,kHz mode of ETMY using a quadrant photodetector on the y-arm transmission port \cite{Blair15, Blair16}. We focus on this mode because it has significant sensing gain. Its frequency is read out through the power spectrum density of the photodetector's signal. Figure~\ref{fig:ModeAmb1008} shows the mode's frequency shift over a 36\,hour lock on 8th October 2015 (blue curve). The figure also shows the ambient temperature data as measured by a sensor inside the vacuum tank. The green curve represents the raw sensor data and the red curve represents the data after a one-pole low pass filter is applied with a time constant of 7.2 hours~\footnote{The time constant relates to the test mass itself only and does not include the surrounding chamber.} determined by our finite element model. We can see about 25\,hours into the lock that the frequency change is completely determined by the ambient temperature fluctuation.

\section{Thermal model: \normalfont{\textit{Heat transfer model of Advanced LIGO test masses}}}

\subsection{Thermal couplings}
  
To improve the accuracy of our thermal modelling, we include the radiative heat transfer between the mirror and elements in proximity of the mirror.

\begin{figure}[!ht]
\begin{center}
\includegraphics[scale=0.6]{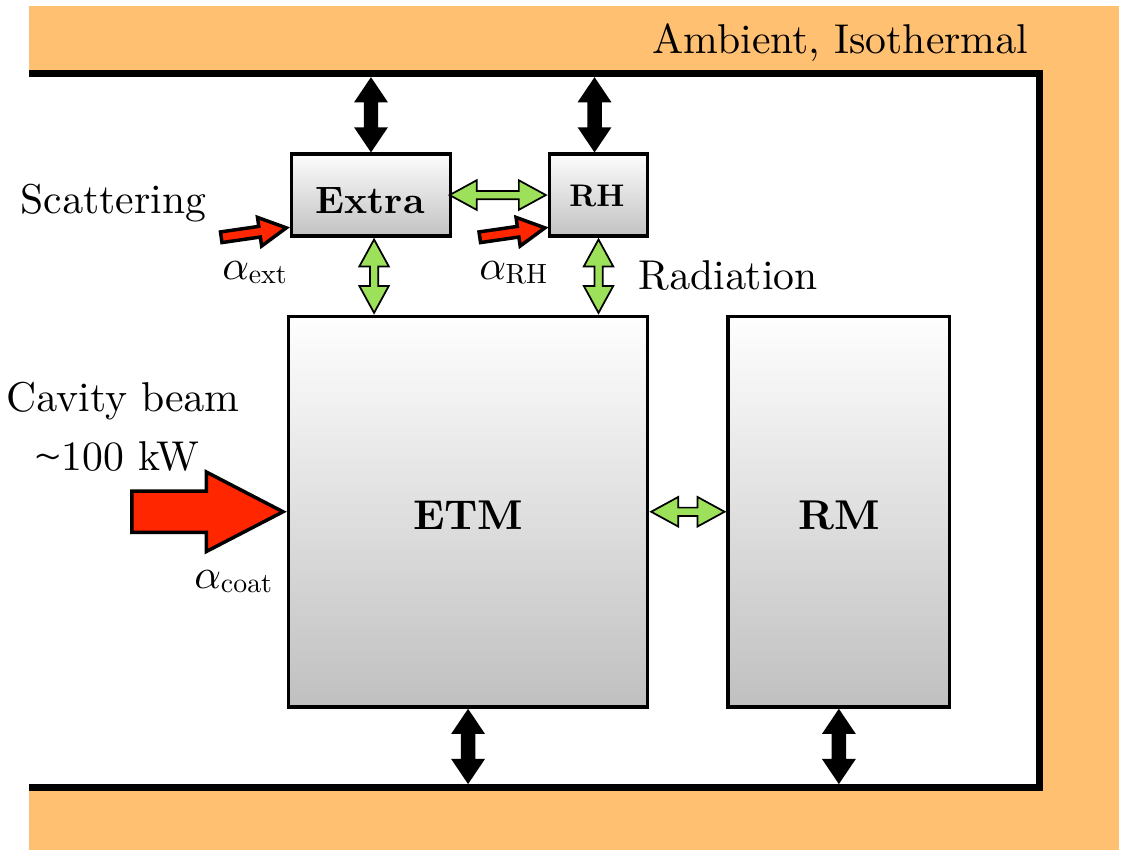}
\caption{A heat transfer model between an end test mass (ETM), a reaction mass (RM), a ring heater (RH), the ambient isotherm, and an extra term representing the complex structures surrounding the test mass that could not be modelled in detail. Red arrows represent energy coming from the intra-cavity beam through coating absorption or scattering. Green arrows represent the radiative heat couplings between different elements and black arrows the radiative heat couplings with the ambient.}
\label{fig:heat_coupling}
\end{center}
\end{figure}


Figure~\ref{fig:heat_coupling} depicts different mechanisms by which heat is transferred in the system. When the arm cavity is locked, $\sim$100\,kW of laser power is incident on the test masses. The absorption, in the order of 1\,ppm, of the high reflectivity coatings results in a portion of the light being converted into heat. In addition, scattered light from the beam illuminates elements surrounding the mirror, part of which will be absorbed and converted into heat by those elements. 

Some of the heat from absorbed scattered light radiatively couples to other elements, introducing longer time constants to the test mass thermal transient. The surrounding elements used in this model are the reaction mass, the ring heater, and an empirical element called the `extra term' that is a simplification of all other nearby objects.

Finally, all elements will exchange heat with the ambient via radiation.  The ambient temperature is recorded by a sensor in the vacuum tank, and the data from this sensor is included in the model. 
The amount of scattered light hitting the RM as well as the radiative coupling between the RM and the RH are negligible and have been omitted from Figure~\ref{fig:heat_coupling} for simplicity.

\subsection{Ambient temperature effect}\label{Ambient temperature effect}

We can treat each of the heat transfer mechanisms discussed in the previous section for the test mass as having a linear effect on its mechanical mode frequency. This constitutes a good approximation since each process only changes the overall temperature of the test mass by approximately 0.2$^{\circ}$C around room temperature. The mechanical mode frequency change of the mirror can thus be expressed as:

\begin{equation}
\Delta f_{m}(t) = \Delta f_{\rm{laser}}(t)+\Delta f_{\rm{surroundings}}(t)+\Delta f_{\rm{ambient}}(t)
\label{eq:modeFReqFunc}
\end{equation}
where the terms on the right hand side correspond to contributions from the laser heating of the ETM, the radiative heat transfer with its surroundings (RH, RM and extra term), and the radiative heat transfer with the ambient respectively.

We examine mode frequency data from 5 locks where the cavity was in a \emph{clean} thermal status. The clean status means that light wasn't circulating in the cavity for at least 30\,hours prior to the lock, and therefore the test mass and surrounding elements were, to a good approximation, in thermal equilibrium. 

For each lock, $\Delta f_{\rm{laser}}(t)$ and $\Delta f_{\rm{surroundings}}(t)$ are calculated by our finite element model while $\Delta f_{\rm{ambient}}(t)$ is not. Instead, it is included by applying the low-pass filter to the measured ambient temperature data as shown in Figure~\ref{fig:ModeAmb1008}.



\section{Finite element model}\label{Finite element model}

We use COMSOL Multiphysics to build a model of the whole thermal system. The heat transfer module is attached to the model to calculate the temperature of the ETM and the ring heater over time. A solid mechanics study is attached to the module calculating eigenfrequencies of the ETM at each time point.

\begin{figure}[!ht]
\begin{center}
\includegraphics[width=12cm]{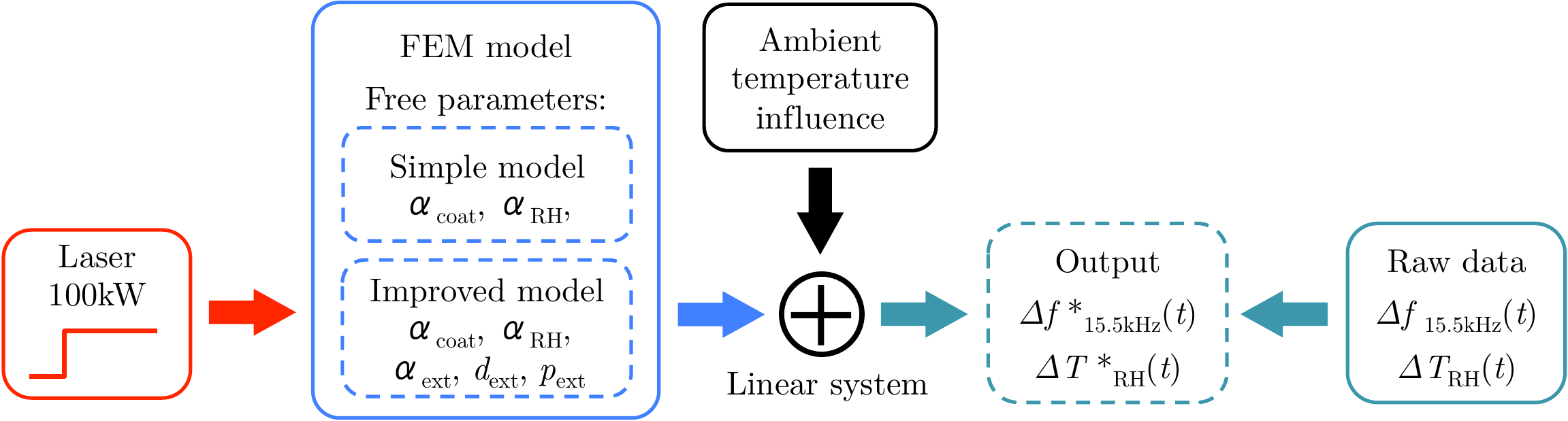}
\caption{Inputs, outputs, and free parameters of the finite-element model.}
\label{fig:fit_parameters}
\end{center}
\end{figure}

The only input of the model is the laser power step function that is 0 when the cavity is unlocked and when the cavity is locked the measured power is typically 100\,kW. The actual time to reach full power can be up to 40 minutes but this has no significant effect on the thermal evolution modelled here. Outputs are the frequency shift of the 15540 Hz mechanical mode and the ring heater temperature change as a function of time. The output data, after adding the ambient temperature influence, is compared with the measured data. There are several free parameters that we can tune to fit our model to the measured data, discussed later in this section. Figure~\ref{fig:fit_parameters} provides an input-output diagram for this model.

\subsection{A simple model}

We start by building a simple model that only considers the end test mass, the reaction mass and the ring heater, because these components have regular shapes and can be easily modelled in COMSOL.

\begin{figure}[!ht]
\begin{center}
\includegraphics[width=5cm]{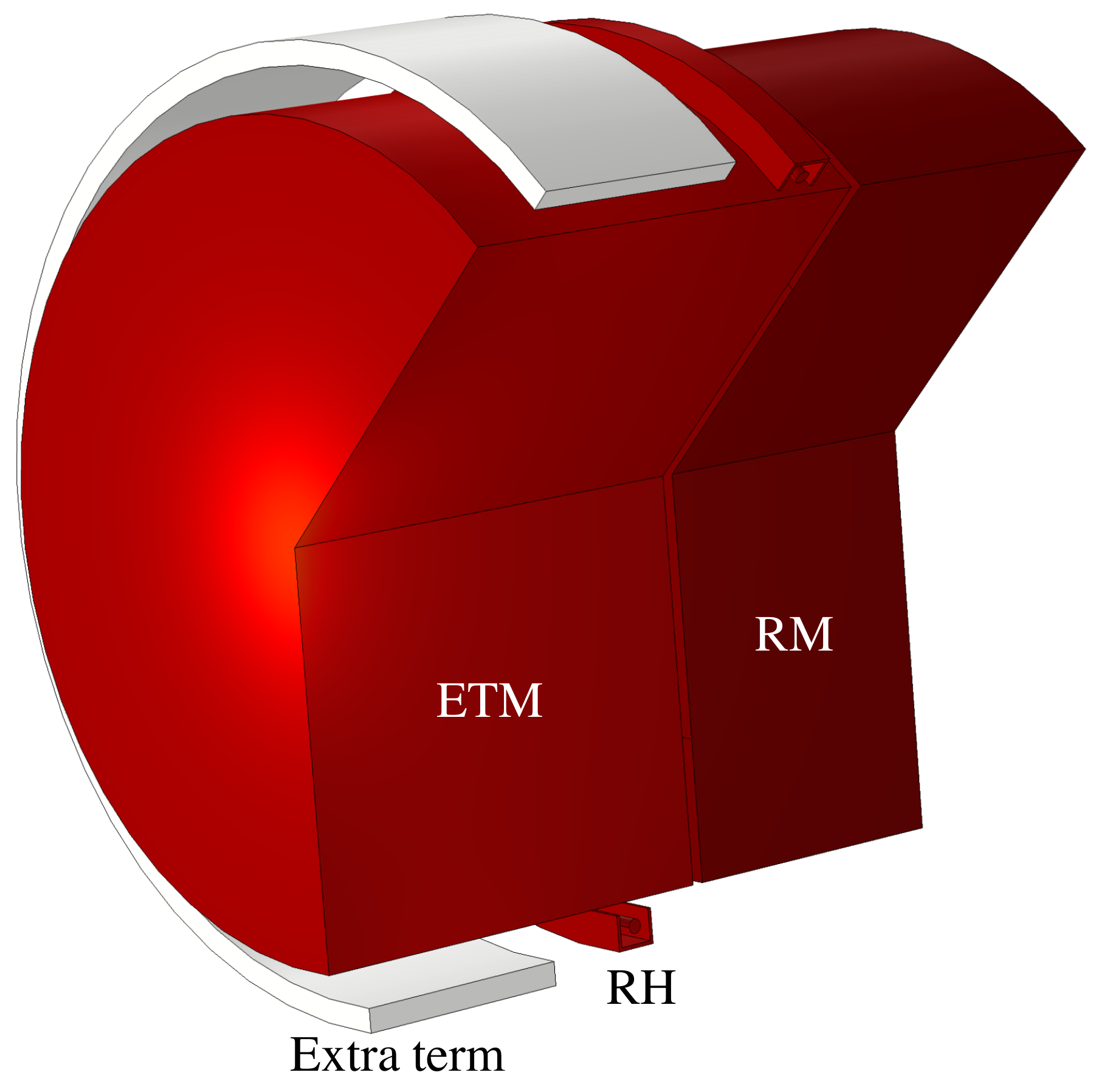}
\caption{\label{fig:3D_configuration} 3D view of the geometry of the ETM, RH, RM, and extra term as modelled in COMSOL (225$^{\circ}$ slice). Geometric parameters of the model are shown in the appendix.}
\end{center}
\end{figure}

\subsubsection{End test mass}

The Advanced LIGO ETM is a cylinder with a 170\,mm radius and 200\,mm thickness made of Heraeus Suprasil 3001 fused silica. When the 100\,kW laser beam impinges on it, the high-reflectivity coating of the ETM absorbs some energy and converts it into heat. In the model, a heat source boundary condition with a Gaussian profile is used to simulate the beam spot on the mirror surface. The intensity of a Gaussian beam with beam radius $w$ and power $P_0$ at a distance $r$ from the beam axis is
\begin{equation}
I(r) = \frac{2P_{0}}{\pi w^{2}} \exp \left( -\frac{2r^{2}}{w^{2}} \right) \hspace{0.2cm} {{\rm W}/{\rm m}^{2}},
\end{equation}
where $w=$6.2\,cm and $P_0$ is the circulating laser power heating the ETM, measured by a photodetector, which is around 100\,kW. The energy absorbed by the coating is given by $\alpha_{\rm{coat}}I(r)$ where $\alpha_{\rm{coat}}$ is the absorption coefficient of the coating. For a laser power of 100\,kW, a coating absorption of 1\,ppm corresponds to a total absorbed energy of 0.1\,W. The high reflectivity and low transmission of the ETM coating result in negligible absorption in the substrate.



\subsubsection{Reaction mass}

The reaction mass (RM) is another fused silica cylinder with the same radius as the ETM, but a thickness of 130\,mm. It is located coaxially 5\,mm away from the rear surface of the ETM. Its primary purpose is to allow longitudinal actuation of the test mass via an electrostatic force.

\subsubsection{Ring heater}

The ring heater (RH) is a glass ring surrounding the ETM that is wrapped with nichrome wire and surrounded by a U-shaped aluminium shield. The inner surface of the aluminium shield is gold-plated to reflect infra-red light. The ring heater is used to create a thermal gradient to compensate for the change in radius of curvature induced by laser heating on the surface of the optic~\cite{Brooks16, Ramette16}. There is a thermometer inside the RH shield monitoring its temperature. During O1 the power fed to the RH was small and fixed (approximately 1\,W), changing the RH temperature by 0.25$^{\circ}$C at around 18$^{\circ}$C from the initial cool status with zero power. Since the ETM-RH system was in thermal equilibrium during our measurements we don't take this gradient into account, as it has a negligible effect on the ETM's temperature fluctuations. The RH is, however, subject to scattering by the ITM and ETM. The thermal sensor inside the RH shield registers a rise in temperature during the first five hours of each lock. In our model this is described by a heat source on the surface of the ring heater shield facing the cavity. The contribution of ambient temperature fluctuations to the RH temperature is determined in a process similar to that in section~\ref{Ambient temperature effect}. The finite element model predicts a time constant for a simple low-pass filter of approximately 1.43 hours.

Material properties of fused silica and aluminium used in the model are listed in Table~\ref{tab:material property}.

\begin{table}[!htp]
\centering
\caption{Material properties}
\label{tab:material property}
\begin{minipage}{12.8cm}
\centering
\begin{tabular}{ccccc}
\toprule
Property & Symbol & Fused silica & Aluminium & Unit \\
\midrule
Density & $\rho$ & 2203 & 2700 & kg/m$^{3}$ \\
Young's modulus at 16$^{\circ}$C & $E$ & 73$\times10^{9}$ & --  & Pa \\
$dE/dT$ at 17$^{\circ}$C~\footnote{This property has been found to be constant over a 40$^{\circ}$C temperature range as $(dE/dT)/E=+1.52\times 10^{-4}/$K~\cite{dEdTE}.} & $dE/dT$ & 11.5$\times10^{6}$ & --  & Pa/K \\
Poisson's ratio & $\nu$ & 0.17 & -- & -- \\
$d\nu/dT$ at 17$^{\circ}$C & $d\nu/dT$ & $5.5\times 10^{-5}$ & -- & $1/$K \\
Thermal conductivity & $k$ & 1.38 & 160 & W/(m$\cdot$K) \\
Thermal expansion coefficient & $\alpha$ & 5.5$\times10^{-7}$ & --  & $1/$K \\
Heat capacity & $C_{p}$ & 740 & 900 & J/(kg$\cdot$K) \\
Relative permittivity & $\epsilon$ & 3.8 & 1 & -- \\
Surface emissivity & $\epsilon_{e}$ & 0.93 & 0.1 & --\\
\bottomrule
\end{tabular}\par
\vspace{-0.5\skip\footins}
\renewcommand{\footnoterule}{}
\end{minipage}
\end{table}

There are only two parameters we can tune in the simple model: the energy absorbed by the ETM coating from the laser beam ($\alpha_{coat}$) and the energy absorbed by the ring heater surface from the scattered light ($\alpha_{RH}$). These two parameters are constrained by the thermal transient in the initial hours of lock, discussed in section~\ref{Tuning the Model}.  After two hours this simple model fails to match the measured data.

\subsection{The improved model}

The short term thermal behaviour of the system is reliably predicted by the simple model. The reliability of long-term models, however, depends on how accurately test mass surroundings are modelled. Simulating all surrounding elements in detail quickly becomes an impractical task as the computation time rises exponentially with each new radiative surface. To approximate effects of all elements in the vicinity of the test mass we improved the simple model by introducing the extra term, an aluminium annulus that surrounds the ETM (see Figure~\ref{fig:3D_configuration}). This extra term couples to the ETM radiatively and also receives scattered light from the cavity. By doing this, we obtain another three free parameters to tune: the energy absorbed by the extra term from the scattered light ($\alpha_{ext}$), and dimensional and position parameters for the extra term ($d_{ext}$ and $p_{ext}$~\footnote{We only chose these two geometric parameters for the extra term as others largely degenerate with them.}), shown in detail in~\ref{app:Geometric parameters}. The method used to tune the improved model are shown in the next section.

\section{Tuning the Model: \normalfont{\textit{Creating an accurate long time-scale model}}}\label{Tuning the Model}

The FEM model calculates eigenfrequencies of the ETM over time and compares results with measured data. Tuning the model required three steps:

1. The laser heating term is estimated from the first three hours of the lock where the dominant term is coating absorption.  This gives an estimate of the coating absorption coefficient, $\alpha_{\rm{coat}}$.

2. The ring heater scattered light heating term is also estimated from the first hours of the lock.  The temperature recorded by the ring heater sensor is used to estimate the amount of scattered light absorbed by this element $\alpha_{\rm{RH}}$.

3. Finally parameters of the extra heating term are obtained.  Setting $\alpha_{\rm{coat}}$ and $\alpha_{\rm{RH}}$ from the previous two steps the extra term parameters are chosen to minimise the residual between the measured data and the long-term simulation of the test mass, including all heat transfer effects depicted in Figure~\ref{fig:heat_coupling}. The parameters adjusted are dimensions, scattered light absorption and the proximity of the ring to the test mass.

\subsection{Coating absorption measurement}\label{Coating absorption measurements}

\begin{figure}[!ht]
\begin{center}
\includegraphics[scale=0.55]{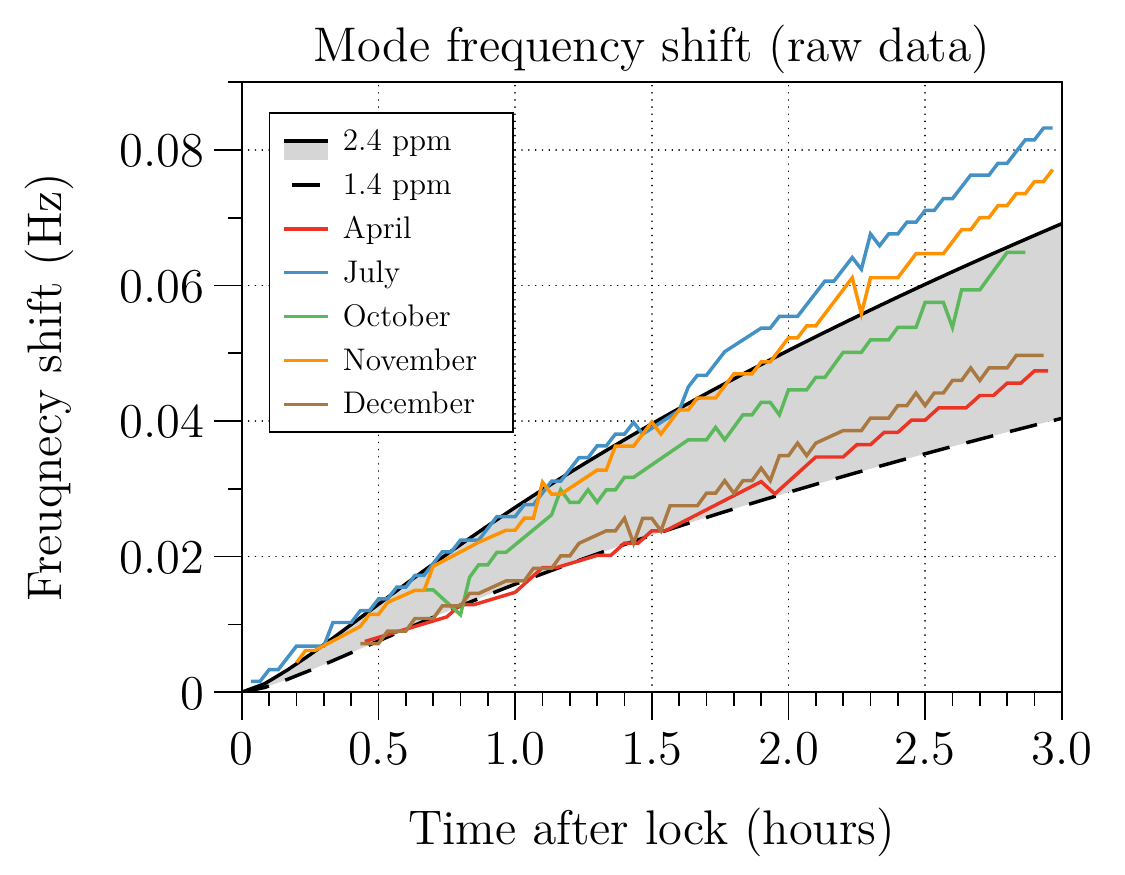} \hspace{0.1cm}
\includegraphics[scale=0.55]{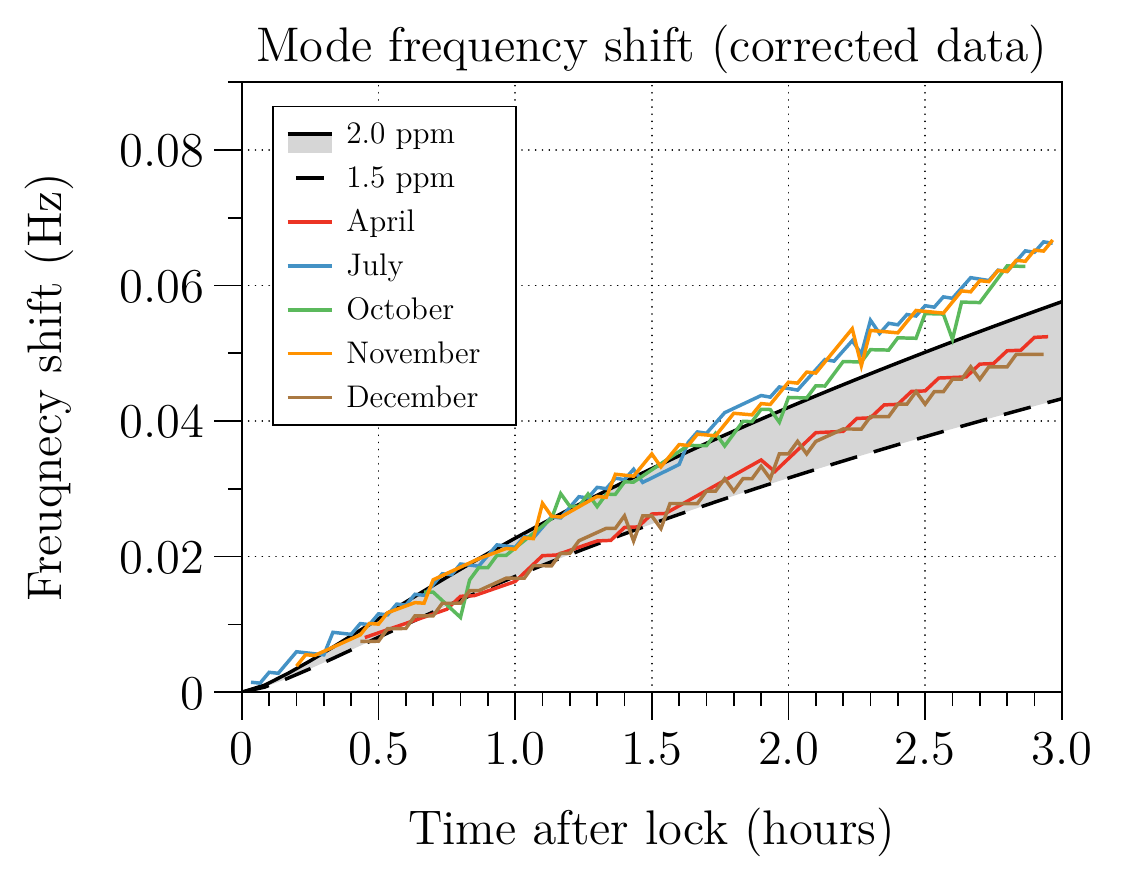}
\caption{Mode frequency shift data from the first 3\,hours after lock. The raw data shows a disagreement because the ambient temperature, the alignment of the arm cavity and the beam heating position in each lock are different. After ambient temperature effects are removed from the data, the trends are in better agreement with modelled coating absorptions of 1.5 to 2.0\,ppm.}
\label{fig:CoAb_ETMY}
\end{center}
\end{figure}

The first three hours of several locks is used to  estimate the energy absorbed by the Livingston ETMY coating. We chose 5 lock periods from April to December 2015 (information of each lock is shown in~\ref{app:start time}). These lock stretches were chosen because in each case there was no light in arm cavities and no anomalies in the laboratory temperature for at least 30\,hours prior. This justifies the assumption that the test mass is in thermal equilibrium with its surroundings at the start of the lock. The raw data from these locks is shown in the left plot of Figure~\ref{fig:CoAb_ETMY}. The circulating power varied from 105\,kW to 112\,kW within the lock segments used here. We have linearly scaled the measured coating absorption to the 100\,kW beam.

The right plot of Figure~\ref{fig:CoAb_ETMY} shows the \textit{corrected data}, where ambient temperature effects are removed from the raw data. The simulated (black) curves display the effect of coating absorption for 1.5 to 2.0\,ppm of the laser power. It is clear that the slow radiative effects start affecting the test mass temperature after about 1.5 hours. All 5 locks display a similar trend but don't exactly overlap. The data seem to bunch in two groups. We suspect this is due to some unmodelled effects of the interferometer. There are many possible explanations for the small differences in the 5 traces, such as error in ambient temperature measurements, the long-term ambient temperature to test mass temperature transfer function (see section~\ref{results}), other heating terms including in-vacuum electronics or difference of the alignment of the arm cavity and the beam heating position in each lock.


\subsection{Scattered Light}\label{Scattered Light}

\begin{figure}[!h]
\begin{center}
\includegraphics[scale=0.55]{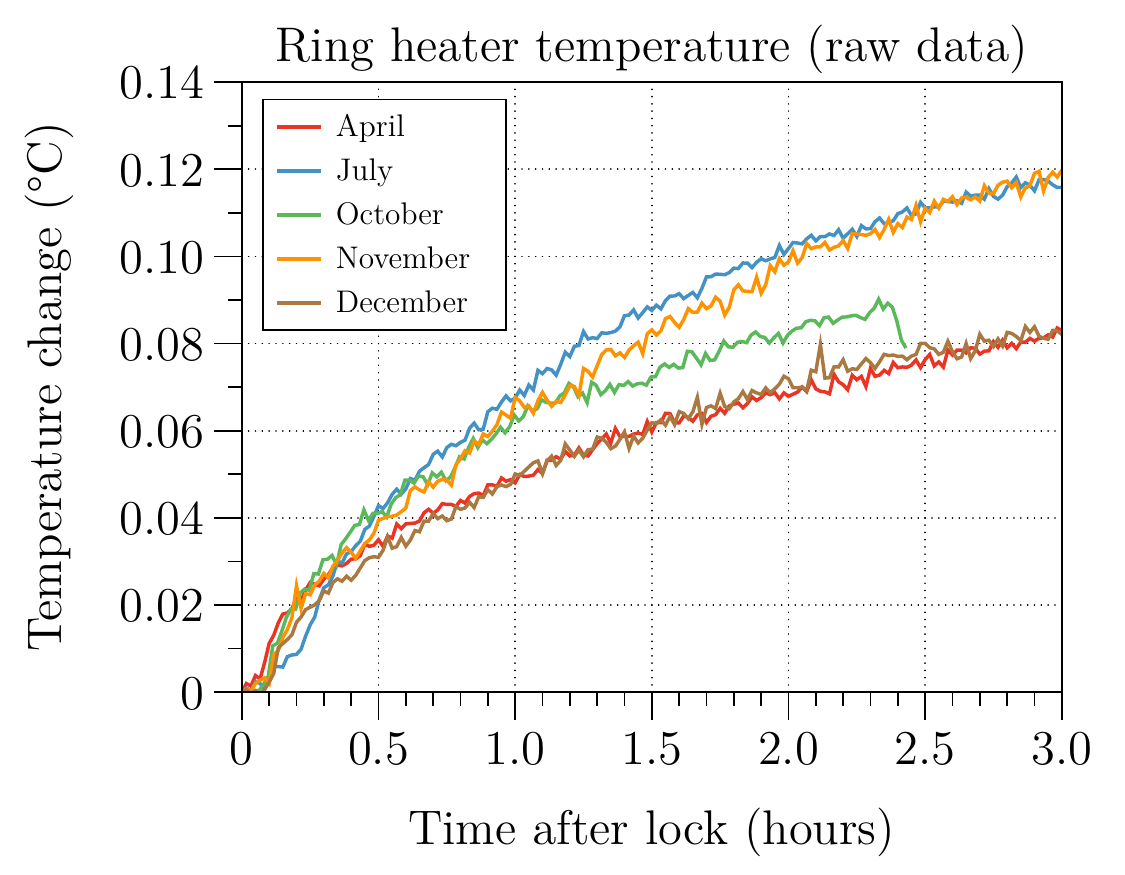}
\includegraphics[scale=0.55]{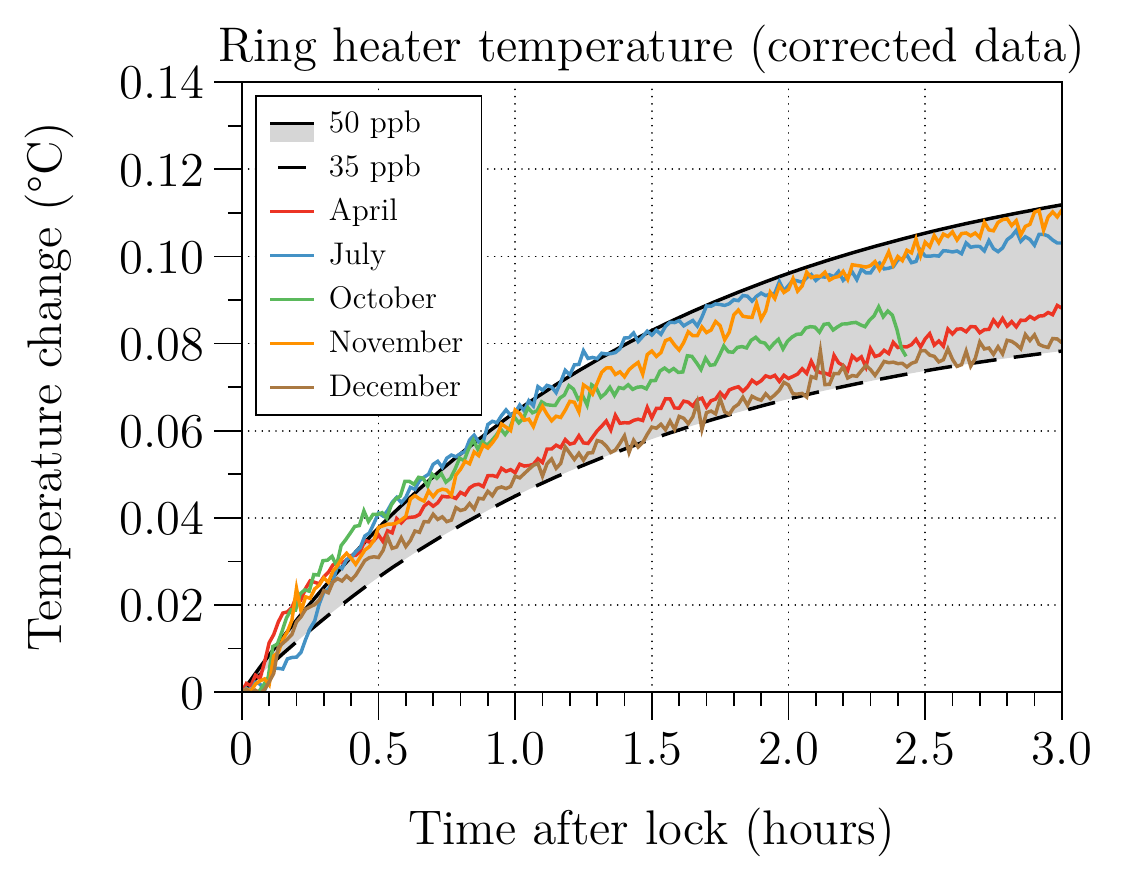}
\caption{Ring heater temperature data from the first 3\,hours after lock. After removing ambient temperature effects, the data agrees with the models for between 35\,ppb and 50\,ppb of ring heater absorption.}
\label{fig:scattering}
\end{center}
\end{figure}

The scattered light absorbed by the ring heater completely dominates the initial thermal transient of this element. The data from the ring heater temperature sensor is compared to the model absorption estimates in Figure~\ref{fig:scattering}.  We obtain estimates for $\alpha_{\rm{RH}}$ of 35 to 50\,ppb of the laser power in the cavity.  The absorption of aluminium at 1064\,nm varies greatly between 4\% and 12\% depending on the surface roughness and the thickness of the oxide layer~\cite{bergstrom08}. As a result, the estimated  scattered light incident on the ring heater could range from 0.3 to 1.25\,ppm of the 100\,kW beam. This energy can come from the small angle scattering of the ITMY or the large angle scattering of the ETMY. For example, the estimated small angle scattering from the ITMY is about 10\,ppm~\cite{Scattering}.

\subsection{Long-term simulation}\label{the long term simulation}

We use the lock period of April 12, 2015, lasting for 31 hours, to tune the parameters of the extra term in the model in this section.

\begin{figure}[!h]
\begin{center}
\includegraphics[scale=0.6]{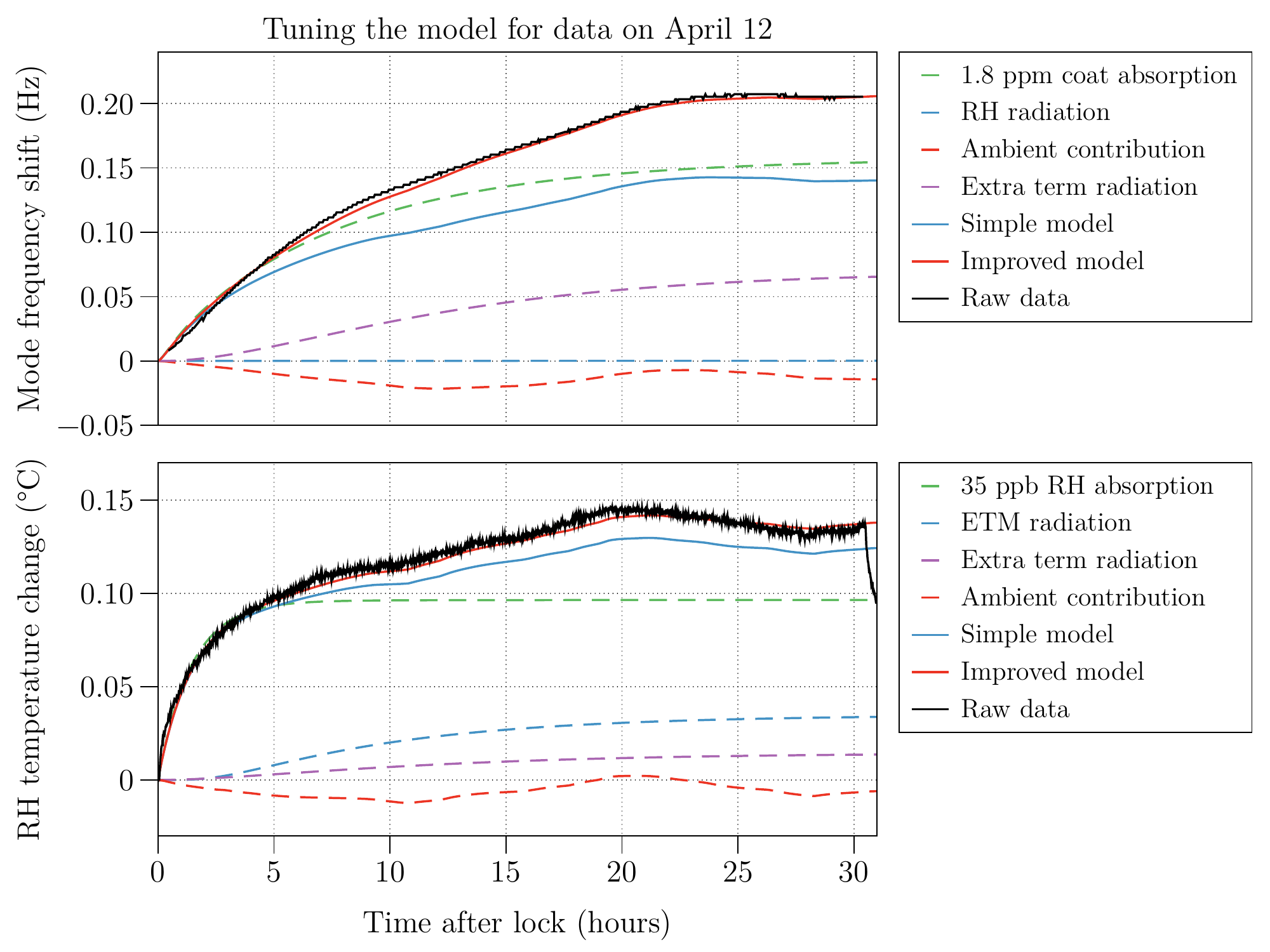}
\caption{Long-term comparison of the April data (black curves) with both models. As temperature changes are small the system is linearised such that the contributions of the various components can be examined separately.
\label{fig:Separated_effect_0412}}
\end{center}
\end{figure}

Our long-term model includes the ETM, RM, RH, and the extra term depicted in Figure~\ref{fig:heat_coupling}. The coating absorption is set to the medium value of the estimated range which is 1.8\,ppm, as derived in section~\ref{Coating absorption measurements}. The scattered light absorbed by the ring heater is chosen to be 35\,ppb of the circulating laser power, because this value agrees better with the April data, as derived in section~\ref{Scattered Light}. Finally the ambient heating term is generated using the similar method shown in Figure ~\ref{fig:ModeAmb1008}.

In Figure~\ref{fig:Separated_effect_0412} a linear decomposition of the contributions to the model are displayed. On the top, the green dashed line shows the coating absorption component of the 15.5kHz mode frequency change. The ring heater scattered light heating component is shown dashed blue, it turns out to be very small. The red dashed line is the ambient temperature contribution to the mode frequency change. The sum of these three contributions determines the simple model shown in solid blue. In the improved model, the parameters of the extra term ($d_{ext}$, $p_{ext}$ and $\alpha_{ext}$) were adjusted based on visual comparison and the resulting error was reduced to less than 5\%. On the bottom the ring heater measured data (black curve) and the component heating terms from our model are shown. The green dashed line is the ring heater scattered light absorption term, the blue dashed line is the radiation influence from the ETM and the dashed red is the ambient contribution. The solid blue and red lines are the sum of component terms corresponding to results of the simple model and improved model respectively.

We can see in both sub-plots that the simple model deviates from the data after about 2 hours. The slow effect indicates radiative coupling, resulting in an maximum error of 40\% in mode frequency change.  
The extra term geometry parameters, $d_{ext}$ and $p_{ext}$, were chosen to match the time constant of the residual based on visual comparison. The extra term absorbed power parameter, $\alpha_{ext}$, was subsequently tuned to minimise the residual between the model output and measured April data; also its value is chosen within the range of the small angle scattering loss from ITMY which is up to 10\,ppm.

\section{Results: \normalfont{\textit{testing the model}}}\label{results}

\begin{figure}[!h]
\begin{center}
\includegraphics[scale=0.6]{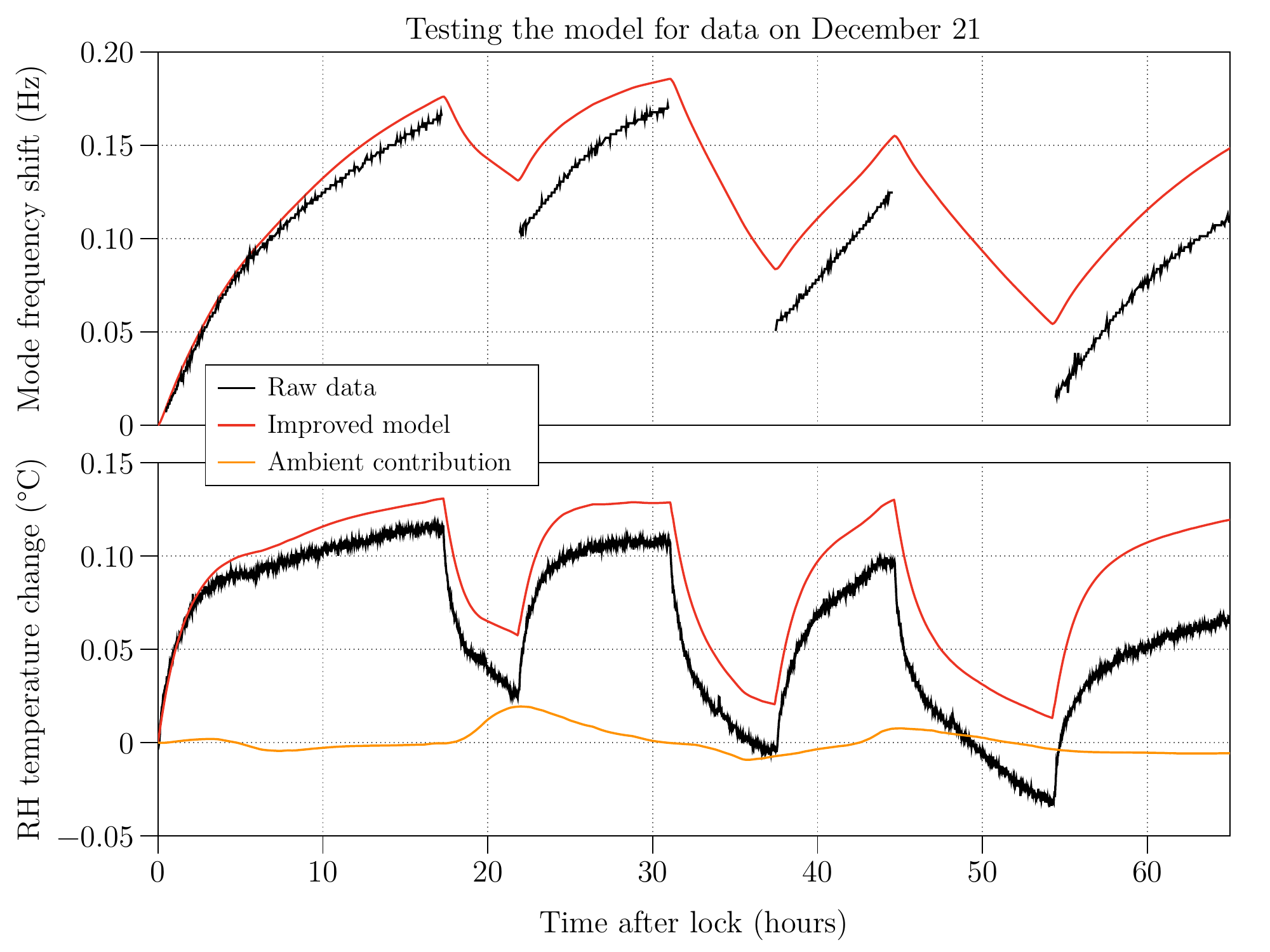}
\caption{Testing the tuned model with the December lock data. The black line is the raw data, the red curve is the improved model, and the orange line in the lower plot represents the ambient contribution to the ring heater temperature.}
\label{fig:Compare_0412_1221}
\end{center}
\end{figure}

Figure~\ref{fig:Compare_0412_1221} shows results of applying the model developed with the April data set to a dataset starting  December 21, 2015. The first lock lasts 17 hours and is followed by 3 locks each of about 10 hours. The black curve is the raw data and the red curve is the prediction from the improved model.

There is good agreement between the model and the measured data in the first lock period, where the deviation between model and experiment is only 3\%. However, there is significant deviation in subsequent locks, caused by different long-term behaviours measured from the ring heater temperature sensor and the ambient temperature sensor. As shown in the lower plot in Figure~\ref{fig:Compare_0412_1221}, the measured ring heater temperature (black curve) at the beginning of the 4th lock is lower than at the start of the 1st lock, where we expect a thermal equilibrium state. This can be only explained by a decrease in ambient temperature. However, the ambient temperature sensor we used displays a flat trend. Data stretches from long locks starting from a clean thermal state are rare, and further tests of this model will require data from LIGOs second observation run.



Despite the possibility that the ambient temperature sensor data is corrupted, we demonstrate a good model estimating the ETMY temperature over the first 17 hour lock. We expect an improvement in the model accuracy given reliable ambient temperature data and following investigations of long-term features of the transfer function from ambient temperature to test mass temperature.

\section{Conclusion and further work}

Mechanical mode frequencies can be used as very accurate thermometers. We have provided an estimate of LIGO Livingston's ETMY coating absorption of 1.5 to 2.0\,ppm. The scattered light incident on the ring heater is estimated to be 0.3 to 1.3\,ppm. We have described a thermal model that estimates the test mass temperature in a long-term scale. The accuracy of the thermal model of the test mass was improved with the addition of an extra term to account for radiative contributions from objects in the vicinity of the test mass. The model is tuned to a data set from April, matching the ETMY measured temperature shift within 5\%. The model is then tested on a data set from December. The agreement is good over the first 17 hour lock but deviates significantly in subsequent locks. We expect to improve the model further when reliable ambient temperature sensors become available.


\section{Acknowledgement}

This work is supported by the LSC fellows program and the Science and Technology Facilities Council (STFC). LIGO was constructed by the California Institute of Technology and Massachusetts Institute of Technology with funding from the National Science Foundation, and operates under Cooperative Agreement No. PHY-0757058. Advanced LIGO was built under Grant No. PHY-0823459. The authors would like to thank Haixing Miao, Daniel Brown, and Anna Green for comments and suggestions. The authors also thank the LIGO Livingston Observatory staff for additional support.

\appendix

\section{Geometric parameters of the model}\label{app:Geometric parameters}

\begin{figure}[!h]
\begin{center}
\includegraphics[scale=0.5]{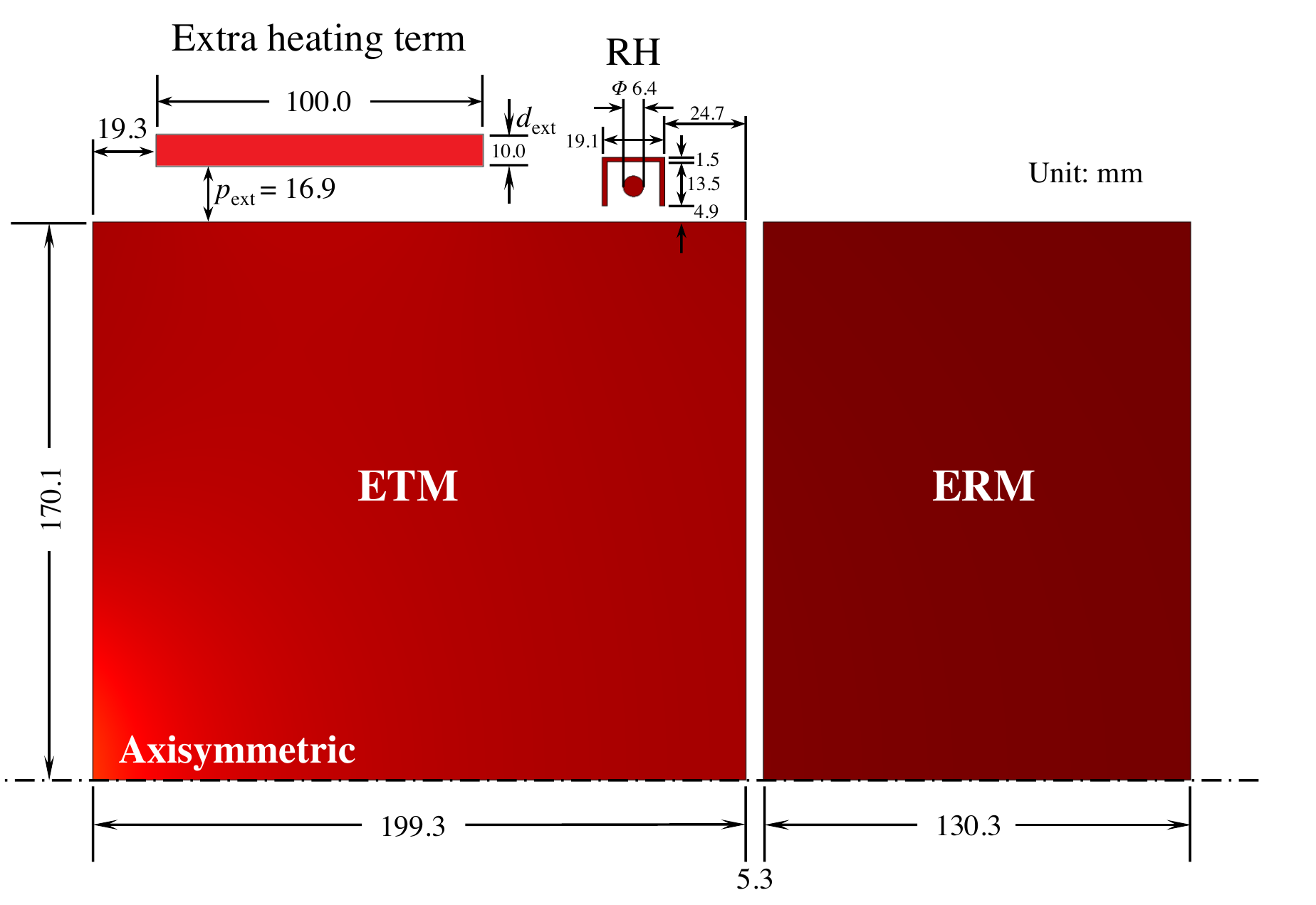}
\caption{Geometric parameters of the model.
\label{fig:2D_configuration}}
\end{center}
\end{figure}

\section{Start time and duration of each lock}\label{app:start time}

\begin{table}[!ht]
\centering
\caption{Start time and duration of each lock of the detector in LIGO, Livingston. The start time is displayed in both UTC (Coordinated Universal Time) and GPS (Global Positioning System) time.}
\label{tab:Start time and duration}
\begin{tabular}{cccc}	\toprule
Date in 2015 & No. & Start time UTC (GPS)  & Duration in seconds (hours) \\	\midrule
12 Apr.     & --  & 02:46:42 (1112842018) & 109799 (30.5)               \\
26 Jul.      & --  & 18:36:13 (1121970990) & 17742 (4.9)                 \\
27 Oct.   & --  & 10:07:18 (1129975655) & 10666 (3.0)                 \\
18 Nov.  & --  & 13:55:52 (1131890169) & 17423 (4.8)                 \\
21 Dec.  & 1   & 21:11:29 (1134767506) & 62368 (17.3)                \\
22 Dec.  & 2   & 19:00:07 (1134846024) & 33347 (9.3)                 \\
23 Dec.  & 3   & 10:38:40 (1134902337) & 26048 (7.2)                 \\
24 Dec.  & 4   & 03:28:02 (1134962899) & 39219 (10.9)
\\	\bottomrule               
\end{tabular}
\end{table}

\section*{References}
\bibliographystyle{myunsrt}
\bibliography{mybib}


\end{document}